# Sociological Inequality and the Second Law

Oded Kafri

Varicom Communications, Tel Aviv 68165 Israel.

## Abstract

There are two fair ways to distribute particles in boxes. The first way is to divide the particles equally between the boxes. The second way, which is calculated here, is to score fairly the particles between the boxes. The obtained power law distribution function yields an uneven distribution of particles in boxes. It is shown that the obtained distribution fits well to sociological phenomena, such as the distribution of votes in polls and the distribution of wealth and Benford's law.

It seems that nature dislike equality. In many cases distributions are uneven, a few have a lot and many have to be satisfied with little. This phenomenon was observed in many sociological systems and has many names. In economy it is called Pareto law [1,2], in Sociology it is called Zipf law [3,4] and in statistics it is called Benford law [5-7]. These distributions differ from the canonic (exponential) distribution by a relatively moderate decay (a power-law decay) of the probabilities of the extremes that enables a finite chance to become very rich. Here it is shown that the power law distributions are a result of standard probabilistic arguments that are needed to solve the statistical problem of how to distribute $P$ particles in $N$ boxes. Intuitively one tends to conclude that $P$ particle will be distributed evenly among $N$ boxes, since the chance of any particle to be in any box is equal, namely, $\frac{1}{N}$.

However, this is an incorrect conclusion, because the odds that each box will score the same amount of particle are very small. Usually there are some lucky boxes and many more unlucky ones. The distribution function of particles in boxes should maximize the entropy. This is because in nature, fairness does not mean an equal number of particles to all boxes $N$, but an equal probability to all the microstates (configurations) $\Omega$. The equal probability of all the microstates is the second law of thermodynamics, which, exactly for this reason, causes heat to flow from a hot place to a cold place.

Calculating the distribution of $P$ particles in $N$ boxes with an equal chance to any configuration is not simple, as the number of the configurations $\Omega(P,N)$ is a function of both $P$ and $N$ namely,

$$\Omega(N,P) = \frac{(N+P-1)!}{(N-1)!P!}. \tag{1}$$

The derivation of the distribution function to Eq.(1) is not new. Planck published it in 1901 in his famous paper in which he deduced that the energy in the radiation mode is quantized [8,9]. Here the Planck's calculation is followed with the modifications needed to fit our, somewhat simpler, problem. Planck first expressed the entropy, namely $S = k_B \ln \Omega$ ($k_B$ is the Boltzmann constant), as a function of the number of modes $N$ and the number of light quanta $P$ in a mode $n = \frac{P}{N}$. Using Stirling formula, he obtained that $S = k_B N\{(1+n)\ln(1+n) - n \ln n\}$. Then he used the Clausius inequality in equilibrium [10] to calculate the temperature $T$, from the expression, $\delta S = \frac{\delta Q}{T} = N \frac{\delta q}{T}$, where $Q$ is the energy of all the radiation modes and $q$ is the energy of a single radiation mode. Therefore, the temperature is $T = N \frac{\partial q}{\partial S}$.

Then, Planck made his assumption that $q = nh\nu$, namely $T = Nh\nu \frac{\partial n}{\partial S}$. Therefore, $\frac{\partial S}{\partial n} = k_B N \ln(\frac{n+1}{n}) = N \frac{h\nu}{T}$, this is the famous Planck equation, namely, the number of quanta in a radiation mode is, $n = \frac{1}{e^{\frac{h\nu}{k_B T}} - 1}$. The calculation of Planck is comprised of three steps. First he expressed the entropy $S$ by the average number of quanta $n$ in a box and the number of boxes (radiation modes) $N$. Next, he used the Clausius equality to calculate the temperature. The equality sign in Clausius inequality expresses the assumption of equilibrium in which all the configurations have the same probability. Then Planck added a new law that was verified by the data of the blackbody radiation that the energy of the quant is proportional to the frequency. This law is responsible for the observation that in the higher frequencies $n$ is lower.

In our problem we do not have energies or frequencies. We just have particles and boxes. Therefore, we will write the dimensionless entropy, namely the Shannon information as a function of $n$ and $N$, and obtain that $I = N\{(1+n)\ln(1+n) - n\ln n\}$. Parallel to Planck, we calculate the dimensionless temperature $\Theta$ according to

$\Theta = \dfrac{\partial P}{\partial I} = N\phi(n)\dfrac{\partial n}{\partial I}$. Here we replace the total energy $Q$ by $P$ and $q$ by $n\phi(n)$, where $\phi(n)$ is a distribution function that tells us the number of boxes having $n$ particles. $\phi(n)$ is the analogue of Planck's $h\nu$. Changing the frequency enabled Planck to change the number of the particles in a mode at a constant temperature. Here we change the probability of a box with $n$ particles at a constant temperature. The sociologic temperature $N\phi(n)\dfrac{\partial n}{\partial I} = \Theta$ is equal, in equilibrium, in all the boxes. Since,

$\dfrac{\partial I}{\partial n} = N\ln(\dfrac{1+n}{n}) = \dfrac{N\phi(n)}{\Theta}$ one obtains that $\phi(n) = \Theta \ln\dfrac{n+1}{n}$. This is the analogue of the Planck's equation, namely $n = \dfrac{1}{e^{\frac{\phi(n)}{\Theta}} - 1}$. When $P$ is large as in many statistical systems, we are interested in the normalized distribution. Since $\sum_{n=1}^{N}\phi(n) = \Theta\ln(N+1)$ we obtain that the normalized distribution function is,

$$\rho(n) = \dfrac{\ln(1+\dfrac{1}{n})}{\ln(N+1)} \qquad (2)$$

This is the main result of this paper. This result can be applied to any natural random distribution of inert particles in $N$ boxes[*].

To check the validity of this distribution we start with Benford's law. Benford's law was found experimentally by Newcomb in the 19[th] century, was extended later by Benford [5] and explained on a statistical basis by Hill [6,7]. It says that in numerical data files, which were not generated by a randomizer, namely balance sheets, logarithmic tables, the stocks value etc, the distribution of the digits follows the equation $\rho(n) = \log(1+\frac{1}{n})$. For example, the frequency of the digit 1 is about 6.5 times higher than that of the digit 9. It is seen that if one substitute in Eq.(2) *N*=9 the Benford law is obtained. One can assume that the digit 1 is a box with *n*=1 particle and *n*=9 is a box with 9 particles. In fact, it is obvious that the equation valid for $n = C \times 1$, for the digit 1 and $n = C \times 9$ for the digit 9, where C is any number bigger than one.

Another way, intriguing even more, to check the informatics Planck distribution of Eq.(2) is to compare its results to polls statistics. In polls there are usually *N* choices and *P* voters that suppose to select their preferred choice. Usually each voter can select only one choice. A poll is not necessarily a statistical system. An example for a non-statistical poll is a poll with the three questions: 1. Do you prefer to be poor? 2. Do you prefer to be young, healthy and rich? 3. Do you prefer to be old and sick? In this poll one expects that most people will vote 2 (at least for themselves). However, it is clear that nobody will make the effort to make this poll, as its result is predictable. However, in the Internet there are many examples of multi-choice votes with unpredictable answers. Here we study three choices polls that were done on the Internet by the Globes newspaper [11] (an Israeli economical daily news) on variety of subjects between 10 Feb. 2008 and 10 Apr. 2008, for eight consecutive weeks on various issues. The results are presented in Fig 1.

| | 1 | 2 | 3 | 4 | 5 | 6 | 7 | 8 | Average | Theoretical |
|---|---|---|---|---|---|---|---|---|---|---|
| A | 55% | 39% | 47% | 64% | 46% | 56% | 65% | 47% | 52% | 50% |
| B | 32% | 38% | 31% | 20% | 37% | 30% | 19% | 33% | 30% | 29% |
| C | 13% | 23% | 22% | 17% | 17% | 15% | 16% | 19% | 18% | 21% |

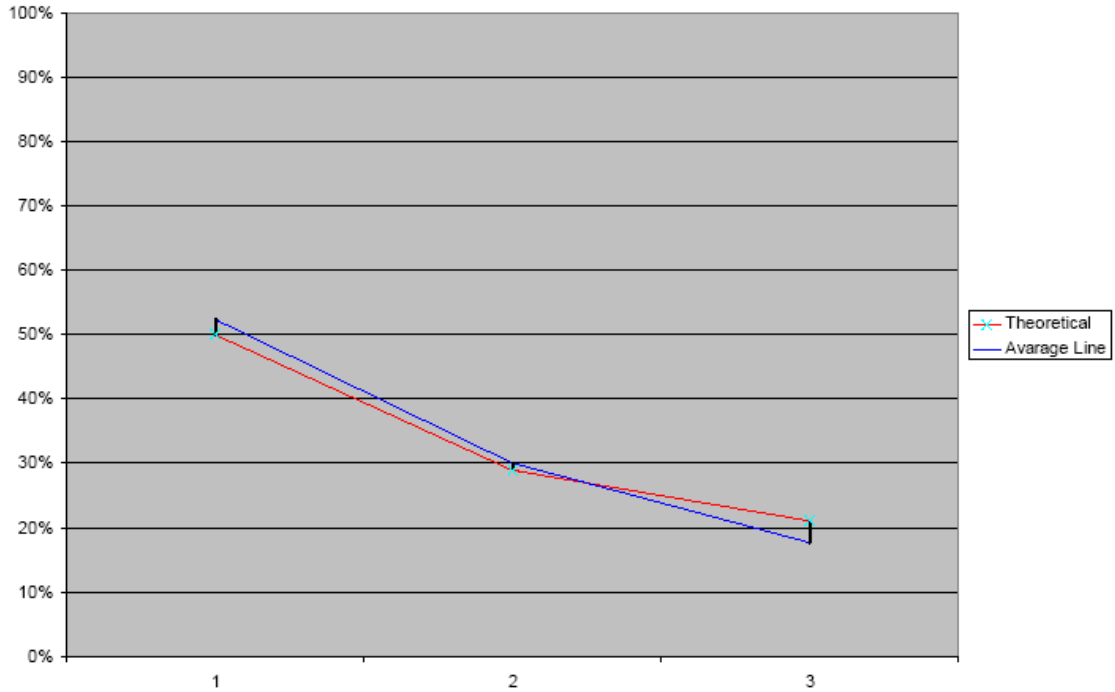

Fig 1. *The average distribution of votes of consecutive eight polls: Each poll has three choices selected by about 1500 voters. The blue line is the actual distribution. The red one is the theoretical calculation based on maximizing the Shannon information.*

It is seen that although the individual votes for the preferred choices A, B and C are quite different from the theoretical values, namely, 50%, 29% and 21% respectively. The average is with a good agreement with the experimental results. It is plausible that on the average, the polls reflect more uncertainty about the best choice than in an individual poll. Therefore, one expects that the average of the eight polls will be closer to equilibrium.

If we consider the number of particles in a box as an indicator of wealth, one may use Eq.(2) to calculate the theoretical particles wealth of boxes in equilibrium. For example, in a set of a million boxes the richest box will have a relative density of $\frac{\ln 2}{\ln 1000001} \cong 0.05$. Namely, 5% of the particles will be in one box. Similarly, the richest 10% will have $\frac{\ln 2}{\ln 11} \cong 0.29$. That means that 10% of the boxes will posses 29% of the particles. The richest half of the boxes will have about 63% of the wealth. The poorest 10% of the boxes will posses $\frac{\ln(1+\frac{1}{9})}{\ln 11} \cong 0.044$ of the particles, namely less than the richest single box. From the point of view of the boxes this is an unfair distribution. Nevertheless, from the point of view of the microstates (which are the configurations of boxes and particles) this is the just way to distribute the wealth.

It was shown previously that Planck formula yields a power law with slop 1[12]. There are many publications that find power-law distributions with variety of slopes [2]. If we assume that the probability of the particles in a box is $\phi^{\alpha}(n)$, we can generalize this theory to a slop $\alpha$ power-law.

To conclude: the uneven distributions that are so common in life are partially an outcome of an unbiased distribution of configurations. This is the second law of thermodynamics as manifested by Boltzmann and Planck. Namely, the probability of all the microstates is equal. Not all the systems are in equilibrium, but systems in equilibrium are more stable. Thermal equilibrium is reached by the dynamics of the system. In blackbody, photons are emitted and absorbed constantly by the hot object, therefore one can expect to a thermal distribution. In economy the money exchanges hands all the time. The digits in numerical data are also changed by the number crunching operations. Nevertheless, the situation in polls is different. Voting in the Internet is a spontaneous non-interactive social activity; therefore, it is surprising that the solitary autonomic action of an individual yields a result of a statistical ensemble.

A possible explanation is that our decision process mimics the behavior of a group, after all a human is a coalition of cells.

\* The Plank derivation can be obtained using a more standard way namely, the Lagrange multipliers. In this method we write a function, $f(n) = \ln\Omega + \beta(P - \sum n\phi(n))$. The first term is the Shannon information and the second term is the conservation of particles. We substitute $\frac{\partial f(n)}{\partial n} = 0$ to find that, $\frac{\beta}{N}\phi(n) = \ln(\frac{n+1}{n})$. This is the maximum information solution that yields after normalization the Eq. (2) see O. Kafri "*Entropy principle in direct derivation of Benford's law*" arxiv: 0901.3047
.